\documentclass[12pt]{article}
\usepackage{times}

\usepackage{amsmath}
\usepackage{amssymb}
\usepackage{float}
\usepackage{graphicx}

\usepackage{color} 

\usepackage{lineno}
\usepackage{url}
\usepackage{pdfpages}
\usepackage{setspace} 

\usepackage{soul}
\usepackage{url}
\usepackage[hidelinks]{hyperref}
\usepackage[utf8]{inputenc}
\usepackage[small]{caption}
\usepackage{amsthm}
\usepackage{booktabs}
\usepackage{algorithm}
\usepackage{algorithmic}
\usepackage[comma,sort&compress]{natbib}
\usepackage{dblfloatfix}

\theoremstyle{definition}

\hypersetup{
	bookmarks=true,         
	unicode=false,          
	pdftoolbar=true,        
	pdfmenubar=true,        
	pdffitwindow=false,     
	pdfstartview={FitH},    
	pdftitle={My title},    
	pdfauthor={Author},     
	pdfsubject={Subject},   
	pdfcreator={Creator},   
	pdfproducer={Producer}, 
	pdfkeywords={keywords}, 
	pdfnewwindow=true,      
	colorlinks=true,       
	linkcolor=blue,          
	citecolor=blue,        
	filecolor=magenta,      
	urlcolor=cyan           
}

\usepackage[labelfont=bf,labelsep=period,justification=raggedright]{caption}

\makeatletter
\renewcommand{\@biblabel}[1]{\quad#1.}
\makeatother

\date{}

\title{Does Spending More Always Ensure Higher Cooperation? An Analysis of Institutional Incentives on Heterogeneous Networks}

\author{
Theodor Cimpeanu$^{1,\star}$,
Francisco C Santos $^{2}$,
and The Anh Han$^{3,\star}$}

\begin{document}

\maketitle
	{\footnotesize
		\noindent
		$^{1}$  School of Mathematics and Statistics, University of St Andrews, United Kingdom\\ 
		$^{2}$  INESC-ID and Instituto Superior Tecnico,
Universidade de Lisboa, Portugal\\ 
     $^{3}$ School Computing, Engineering and Digital Technologies, Teesside University\\ \\
		$^\star$ Corresponding authors: Theodor Cimpeanu (tic1@st-andrews.ac.uk),  The Anh Han (T.Han@tees.ac.uk)
	}

\newpage
\section*{Abstract}
Humans have developed considerable machinery used at scale to create policies and to distribute incentives, yet we are forever seeking ways in which to improve upon these, our institutions. Especially when funding is limited, it is imperative to optimise spending without sacrificing positive outcomes, a challenge which has often been approached within several areas of social, life and engineering sciences. These studies often neglect the availability of information, cost restraints, or the underlying complex network structures, which define real-world populations. Here, we have extended these models, including the aforementioned concerns, but also tested the robustness of their findings to stochastic social learning paradigms. Akin to real-world decisions on how best to distribute endowments, we study several incentive schemes, which consider information about the overall population, local neighbourhoods, or the level of influence which a cooperative node has in the network, selectively rewarding cooperative behaviour if certain criteria are met. Following a transition towards a more realistic network setting  and stochastic behavioural update rule, we found that carelessly promoting cooperators can often lead to their downfall in socially diverse settings. These emergent cyclic patterns not only damage cooperation, but also decimate the budgets of external investors. Our findings highlight the complexity of designing effective and cogent investment policies in socially diverse populations.

 \vspace{0.2in}
 
 \noindent \textbf{Keywords:} Evolutionary game theory, evolution of cooperation, cost-efficiency, incentives, scale free networks, prisoners' dilemma.
 
 \newpage


\section{Introduction}

The design of mechanisms that encourage pro-social behaviours in populations of self-regarding agents is recognised as a major theoretical challenge within several areas of social, life and engineering sciences. It is ubiquitous in real-world situations, not least ecosystems, human organisations, technological innovations and social networks \citep{santos2006pnas,Sigmund2001PNAS,raghunandan2012sustaining,han2019modelling,Andras2018TrustingSystems,dafoe2021cooperative}.  
In this context, cooperation is typically assumed to emerge from the combined actions of individuals within the system. However, in many scenarios, such behaviours are advocated and promoted by an external party, which  is not part of the system, calling for a new set of heuristics capable of {\it engineering} a desired collective behaviour in a self-organised complex  system \citep{penn2010systems}. Among these heuristics, several have been identified as capable of promoting desired behaviours at a minimal cost \citep{chen2015first,han2018ijcai, han2018cost,DuongHanPROCsA2021, cimpeanu2021cost,wang2019exploring,CIMPEANU2023113051,duong2022cost}. However, these studies neglect the diversified nature of contexts and social structures which define real-world populations. Here, we analyse the impact of diversity by means of scale-free interaction networks with dissimilar levels of clustering \citep{barabasi1999emergence,santos:2005:prl}, and test various interference mechanisms using simulations of agents facing a cooperative dilemma.

For instance, if one considers a near future, where hybrid societies comprising humans and machines shall prevail, it is important to identify the most effective incentives to be included to leveraging cooperation in such hybrid collectives \citep{paiva2018engineering,dafoe2021cooperative,Andras2018TrustingSystems,han2021trust}. In a different context, let us consider a wildlife management organisation (e.g., the WWF) that aims  to maintain a desired level of biodiversity in a particular region. In order to do that, the organisation, not being  part of the region's ecosystem, has to decide whether to modify the current population of some species, and if so, then when, and in what degree to \emph{interfere} in the ecosystem (i.e., to modify the composition of the population) \citep{levin2000multiple}. Since a more impactful intervention typically implies larger costs in terms of human resources and equipment, the organisation has to achieve a balance between cogent wildlife management and a low total investment cost. Moreover, due to the evolutionary dynamics of the eco-system (e.g., frequency and structure dependence) \citep{santos2006pnas,key:Hofbauer1998,maynard-smith:1982to}, undesired behaviours can reoccur over time, for example when the interference was not sufficiently strong in the past. 
Given this, the decision-maker also has to take into account the fact that it will have to repeatedly interfere in the eco-system in order to sustain levels of biodiversity over time.
That is, they must find an efficient interference mechanism that leads to their desired goals, while also keeping in mind potential budget concerns.

Specifically, we consider populations of  individuals distributed in a scale-free network, who interact  with their neighbours via the one-shot Prisoner's Dilemma (PD),  where uncooperative behaviour is preferred over cooperation \citep{key:Sigmund_selfishness,santos2006pnas,perc2017statistical,rand2013cooperation}.
As an outside decision maker, we aim to promote cooperation by interfering in the system,  rewarding particular agents in the population at specific moments. The research question here is to identify when and how much to invest (in individuals distributed in a network) at each time step, in order to achieve cooperation within the system such that the total cost of interference is minimised, taking into account the fact that individuals might have different levels of social  connectivity. For instance, we might wonder whether it is sufficient to focus the investment only on highly connected cooperators, as they are more influential. Would targeting influencers reduce overall costs? Do we need to take into account a neighbourhood's cooperativeness level, which was shown to play an important role in square lattice networks \citep{han2018ijcai}? Also, when local information is not available and only global statistics can be used in the decision making process, how different are the results  in heterogeneous networks, in comparison to regular graphs (i.e. homogeneous networks)?  

To answer these questions, this paper will systematically investigate different general classes or approaches of interference mechanisms, which are based on {\it i)}  the global population statistics such as its current composition, {\it ii)} a node's degree centrality in the network and {\it iii)} the neighbourhood properties, such as local cooperativeness level.

Our results show that interference on scale-free networks is not trivial. In particular, we highlight that the inconsiderate distribution of incentives can lead to the exploitation of cooperators. We present which mechanisms are more efficient at fostering cooperation, arguing that social diversity and the network's clustering coefficient both play a key role in the choice of interference mechanisms available to institutions wishing to promote cooperation. 


\section{Model and Methods}
\label{models-methods}

\subsection{Prisoner's Dilemma on Scale Free Networks }
\label{pd}

We consider a population of agents on scale-free networks of contacts (SF NoCs) --- a  widely adopted heterogeneous population structure in population dynamics and evolutionary games. We focus our analysis on the efficiency of various interference mechanisms in spatial settings, adopting an agent-based model directly comparable with the setup of recent lab experiments on cooperation \citep{rand2014static}. Moreover, we select an initial number of nodes $m_0 = 2$, with two additional edges being created at every time step of network generation. This produces networks of average connectivity $z = 4$, serving as a direct comparison between this work and other studies performed on structured populations \citep{han2018ijcai}.

Initially each agent in a population of size $N$ is designated either as a cooperator (C) or defector (D) with equal probability. Agents' interaction is modelled  using the one-shot Prisoner's Dilemma game, where mutual cooperation (mutual defection) yields the reward $R$ (penalty $P$) and unilateral cooperation gives the cooperator  the sucker's payoff $S$ and the defector  the temptation $T$. As a popular interaction model of structured populations \citep{szabo2007evolutionary}, we adopt the following scaled payoff matrix of the PD (for row player): 
{\[
\bordermatrix{~ & C & D\cr
                 C & 1 & 0 \cr
                 D & b & 0  \cr
                },\]}
 with $b$ ($1 < b \leq 2$) representing the temptation to defect. We adopt this weak version of the Prisoner's Dilemma in spite of cooperation prevalence shown in previous works on scale-free networks  \citep{santos2008social}, so as to have a direct  comparison with studies on the effects of rewarding mechanisms in different types of networks \citep{han2018ijcai}.

At each time step or generation, each agent plays the PD with its immediate  neighbours. The score for each agent is the sum of the payoffs in these encounters. Before the start of the next generation, the conditions of interference are checked for each agent and, if they qualify, the external decision maker increases their payoff. Multiple mechanisms (i.e. multiple conditions) can be active at once, but the individual investment cannot be applied more than once; the schemes determine the eligibility for investment. 

At the start of the next generation, each agent's strategy is updated using one of two social learning paradigms -- a \textit{deterministic}, or a \textit{stochastic} rule. Using a deterministic update rule, each agent will choose to imitate the strategy of its highest scored neighbour \citep{nowak1992evolutionary,szabo2007evolutionary}. In the stochastic case, instead of copying the highest scored neighbour, at the end of each generation an agent $A$ with score $f_A$ chooses to copy the strategy of a randomly selected neighbour agent $B$ with score $f_B$ with a probability given by the Fermi rule \citep{traulsen2006stochastic}: $$(1+e^{(f_A - f_B)/K})^{-1},$$ where $K$ denotes the amplitude of noise in the imitation process \citep{szabo2007evolutionary}. In line with previous works and lab experiments \citep{szabo2007evolutionary,rand2013evolution,zisisSciRep2015}, we set $K = 0.1$ in our simulations. Our analysis will be based on this standard evolutionary process  in order to focus on understanding the cost-efficiency of different interference mechanisms. 

We simulate this evolutionary process until a stationary state or a cyclic pattern is reached. The simulations converge quickly in the case of deterministic update, with the exception of some cyclic patterns which never reach a stationary state. Because this work studies cost effective intervention, these rarely-occurring patterns which inherently invite very large total costs are escaped early by running simulations for 75 generations (deterministic update) and 500 generations (stochastic update), at which point the accumulated costs are excessive enough for this mechanism to not be of interest. The difference in the final number of generations accounts for the slower convergence  time associated with stochastic dynamics. Moreover, the results are averaged for the last 25 generations of the simulations for a clear and fair comparison (e.g. due to cyclic patterns). In order to improve accuracy related to the randomness of network topology in scale-free networks, each set of parameter values is ran on 10 different pre-seeded graphs for both types of SF NOCs. Furthermore, the results for each combination of network and parameter values are obtained from averaging 30 independent realisations. It is important to note that the distribution of cooperators and defectors on the network is different for every realisation. 

Note that we do not consider mutations or random explorations when employing a determinstic update rule. Thus, whenever the population reaches a homogeneous state (i.e. when the population consists of 100\% of agents adopting the same strategy), it will remain in that state regardless of interference. Hence, whenever detecting such a state, no further interference will be made. Errors can sometimes occur under the presence of stochastic imitation, thus we never preemptively pause these simulations. Given the difference in convergence time, network size and stopping conditions, we do not directly compare the total costs between these two paradigms.  

\subsection{Cost-Efficient Interference in Networks}

We aim to study how one can efficiently interfere in spatially heterogeneous populations to achieve high levels of cooperation while minimising the cost of interference. An investment decision consists of a cost  $\theta > 0$ to the external decision-making agent/investor, and this value $\theta$ is added as surplus to the payoff of each suitable candidate. In order to determine cost-efficiency, we vary $\theta$ for each proposed interference strategy, measuring the total accumulated costs to the investor. Thus, the most efficient interference schemes will be the ones with the lowest relative total cost.

Moreover, in line with previous works on network interference \citep{chen2015first,han2018ijcai,han2018cost},  we compare  global  interference strategies where investments are triggered based on network-wide information, local neighbourhood information, and, lastly, node centrality information. 

In the \textit{population-based} (\textbf{POP}) approach, a decision to invest in desirable behaviours is based on the current composition of the population. We denote $x_c$ the  fraction  of individuals  in the population adopting cooperative behaviour. Namely, an investment is made if $x_c$ is  at most equal to a threshold $p_c$ (i.e. when $x_c \leq p_c$), for $0 \leq p_c \leq 1$. They do not invest otherwise (i.e. $x_c > p_c$).  The value $p_c$ describes how rare the desirable behaviours should be to trigger external support.  

In the \textit{neighbourhood-based} (\textbf{NEB}) approach, committing an abuse of notation, a decision to invest is based on the fraction $x_c$ of neighbours of a focal individual with the desirable behaviours, calculated at the local level. Investment happens if $x_c$ is  at most equal to  a threshold $n_c$ (i.e. when $x_c \leq n_c$), for $0 \leq n_c \leq 1$; otherwise, no investment is made.

As the presence of structural heterogeneity  in scale-free networks introduces a level of inequality between nodes in terms of influence, we also examine a \textit{node-influence-based} (\textbf{NI}) approach. 
To achieve this, we make use of degree centrality. We denote by $x_i$ the node's centrality measure. The decision-maker invests in a cooperator node $C$ when the value of its degree centrality is above a threshold of influence $c_I$, for $0 \leq c_I \leq 1$. Otherwise, i.e. when $0 \leq x_i < c_I$, no investment is made. The value $c_I$ describes how influential a cooperator node should be to trigger an investment into its survival.

For the POP and NEB schemes, the threshold signifies an increase in the number of nodes that satisfy the requirements for investment. In other words, a threshold of 1 means always investing in all nodes which follow the desired strategy. Conversely, a lower threshold implies a more careful approach to investment, whereby the exogenous agent is stricter in their selection of suitable candidates. The opposite is true for NI, as a value of 1 implies only the most connected individual(s) is eligible for investment; whereas a value of 0 means investing in every cooperative agent.

Interestingly, we posit that these mechanisms require different levels of information, which may or may not be readily available in the given network. In some cases, such as social networks, the connectivity (i.e. the number of friends) of a node is virtually free information which requires no effort on the part of the external decision maker to discern. On the other hand, neighbour-hood based approaches inherently require more information about the population and the level of cooperativeness in different parts of the network. Thus, POP is a broad mechanism which only requires knowledge about overall cooperativeness, but NEB invites complex information gathering, in order to determine the cooperativeness in each neighbourhood. Combining NI with NEB does not require any additional observation than NEB by itself. Our study of neighbourhood based interference does not directly take into account the cost of gathering information, it is a comparison between perceived gains in cooperation and the associated per-individual cost of interference set out in the interference mechanisms. Our discussion will naturally present these subtle differences in the hierarchy of information gathering, as they signal hidden costs for some application domains.

\section{Results}

In contrast to the study on square lattice networks~\citep{han2018ijcai}, we found that performing cost-effective interventions on SF NOCs exhibits complex patterns and presents multiple concerns. In structured populations, more detailed observations resulted in effective interventions with improved outcomes. On the other hand, more knowledge about the population in SF NOCs simply reduces the risk of interfering to the detriment of cooperators. In other words, interfering in SF NOCs without adequate knowledge should be approached cautiously or it could act to the benefit of defectors. This issue is prevalent in SF networks with low clustering (BA model), but also sees some representation in highly clustered (DMS) networks if stochastic dynamics are taken into account. 

Successfully investing in BA populations broadly requires heavy-handed investment and large individual endowments (often orders of magnitude higher than similar mechanisms performed on square lattice populations) or a blanketing mechanism that targets all or almost all cooperators, even those which are not necessarily in danger of converting to defection. Converging to 100\% C is very difficult unless both of these conditions are met and this introduces multiple concerns in the role of an exogenous interfering party. We avoid focusing on solutions where the per-generation cost is excessive, as it is unlikely for any institution to be able to produce unrealistically high endowments, as required by these heterogeneous networks. Instead we focus on effective intervention with manageable amounts of per-generation cost. In the following subsections, we structure our results based upon the most important findings, and provide relevant references to each studied investment scheme where appropriate. Initially, we will present the results for the deterministic update scheme, then the stochastic update, pointing out any difference between the two. All the main findings are robust irrespective of the social learning paradigm employed.


\begin{figure}
\centering
\includegraphics[width=1\linewidth]{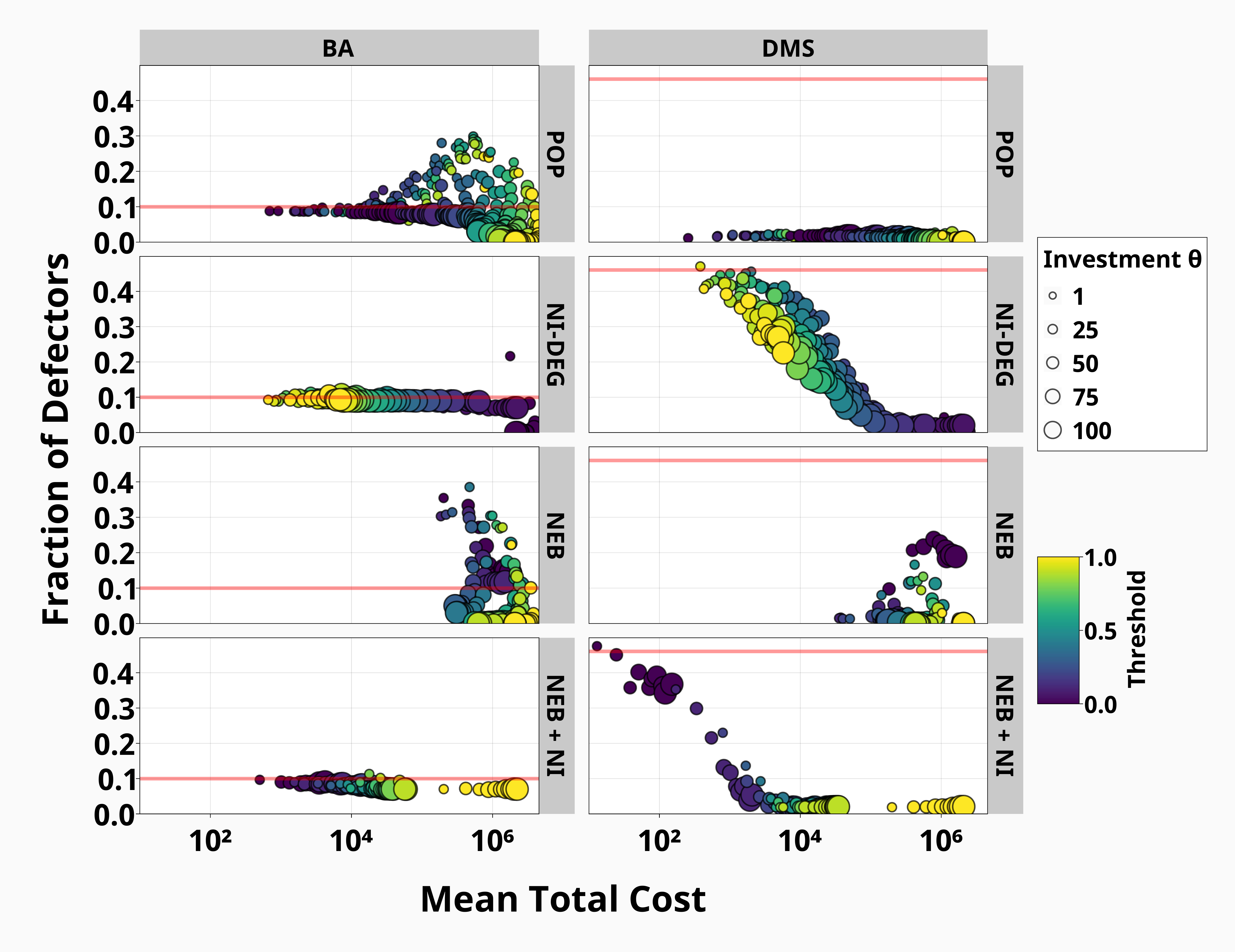}
\caption[Fraction of defectors as a function of the mean total cost for each scheme (deterministic update)]{\textbf{Fraction of defectors as a function of the mean total cost for each scheme (deterministic update)}. The markers' size is determined by the individual investment $\theta$ (grouped to the nearest value), whereas the colour indicates the threshold. Points near the origin indicate the optimal solutions. The horizontal red lines indicate the baseline level of defection in the absence of rewards for either network type (i.e. BA or DMS). Parameters: $b = 1.8; \ N = 5000$.}
\label{fig:pd-pareto-det}
\end{figure}

\subsection{Careless rewarding leads to the exploitation of cooperators} \label{badrewards}

\begin{figure}
\centering
\includegraphics[width=\linewidth]{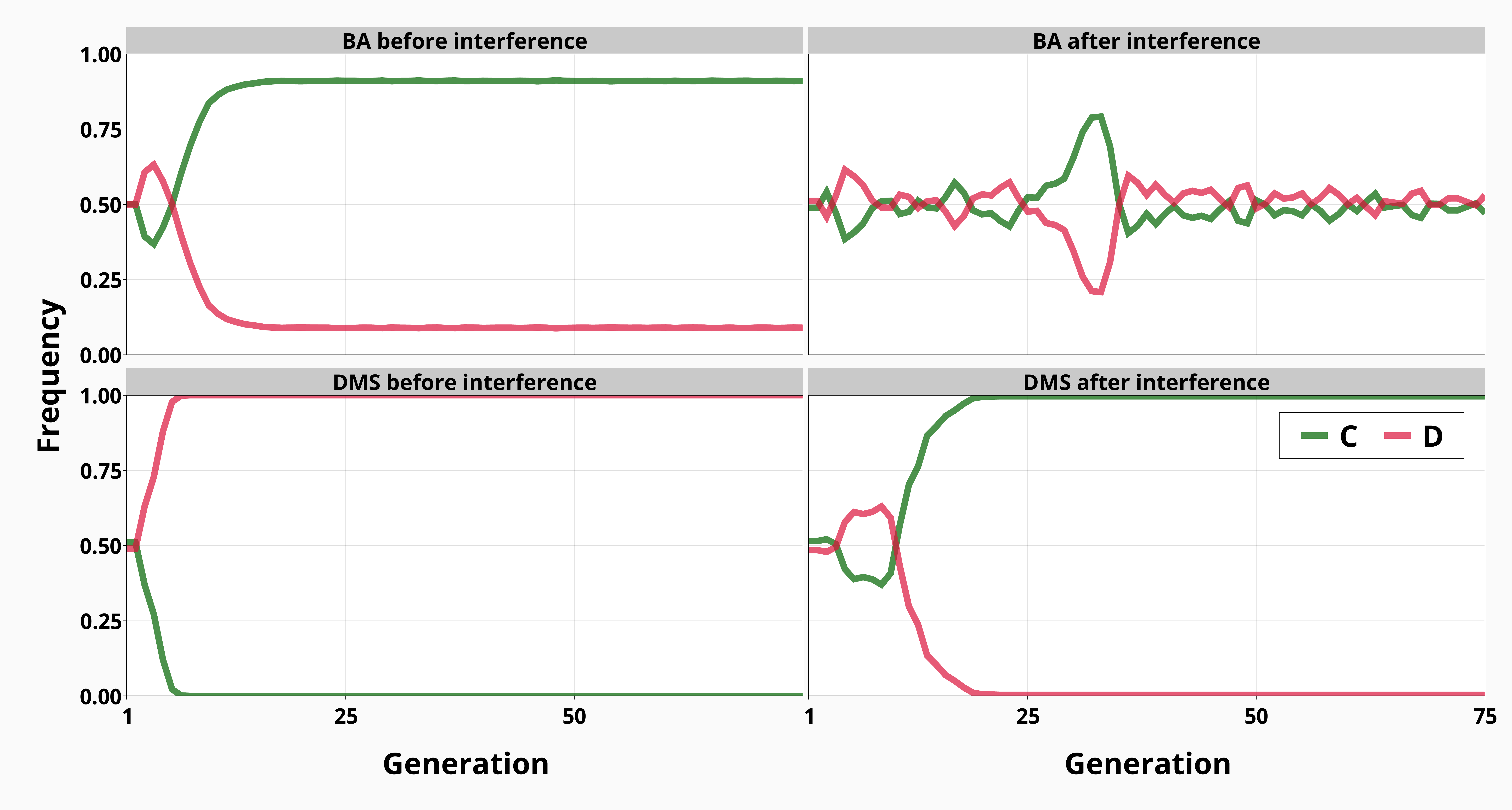}
\caption[Typical time-evolution of cooperation]{\textbf{Typical time-evolution of cooperation,} for $\theta = 5, p_C = 0.8$ (deterministic update). The left column shows the network without interference, while the right one shows the same network after population-based (POP) interference. Some configurations for BA resolve to full C, here we show the scenario in which they do not. Other parameters: $b = 1.8; \ N = 5000$.}
\label{fig:pd-evol}
\end{figure}

In direct contrast with previous findings for positive incentives \citep{han2018ijcai, han2018cost, chen2015first}, an external decision maker should only interfere in scale-free networks with great care, as investing indiscriminately can lead to the detriment of cooperation (see Figure \ref{fig:pd-pareto-det}). We observe that \textit{inclusive} approaches to interference negatively impact the mean frequency of cooperation if the individual endowments are not sufficient to turn defectors away from the temptation of defecting. By inclusive approaches, we imply high values for the threshold that determines the eligibility of investment (for POP and NEB schemes). If an external investor \textit{hedges their bets}, targeting a wide spread of nodes (high threshold) with reduced individual endowments, they risk dooming cooperators. In such a scenario, we see the formation of cyclic patterns, ultimately allowing  D players to exploit cooperators (see Figure \ref{fig:pd-evol}). In this way, an investor would be artificially allowing the survival of cooperators in clusters dominated by defectors, abetting the possibility of these sparsely connected clusters to take over larger formations which cannot easily be maintained by defectors. We note that some of these cyclic patterns eventually converge to a stable state, but the accumulated costs of interference at the end of these long-lasting patterns is prohibitively large. 

In the presence of deterministic selection, this finding is mostly restricted to classical scale-free networks with low clustering (generated using the BA model), but relaxing the intensity of selection produces similar results even with more realistic levels of clustering (see Figure \ref{fig:pd-pareto-stochastic}). Social diversity changes the inherent nature of the problem of rewarding cooperators effectively. Previous results show the emergence of cooperation in heterogeneous networks \citep{santos2008social,santos2006pnas} (shown also in horizontal red lines in Figure \ref{fig:pd-pareto-det}). Compared to homogeneous (well-mixed) and structured populations, there is little improvement to be made in these settings. As the room for improvement narrows, the risk of acting to the detriment of cooperators increases. Individual benefactors prosper temporarily, but the recipients of their naivety are none other than the defectors who exploit them.

\subsection{Clustering reduces the burden of investment}

Real-world networks have been observed to have higher levels of clustering than what normally occurs in typical scale-free networks \citep{Su2016,barrat2005rate}. Nevertheless, several domains, such as the topology of the WWW remain, in which the nodes are sparsely clustered \citep{Albert1999,barabasi2014linked, Barabasi2016}. Thus, it is important to design interference schemes which can target either type of scale-free networks, especially so if there exists a degree of uncertainty about the presence of clusters, or if measuring this factor is unfeasible. We have already mentioned the risks associated with inadequate reward mechanisms, but now we can turn to unveiling the benefits associated with social diversity and clustering in the quest towards engineering pro-social behaviour.

\begin{figure}[t]
\centering
\includegraphics[width=\linewidth]{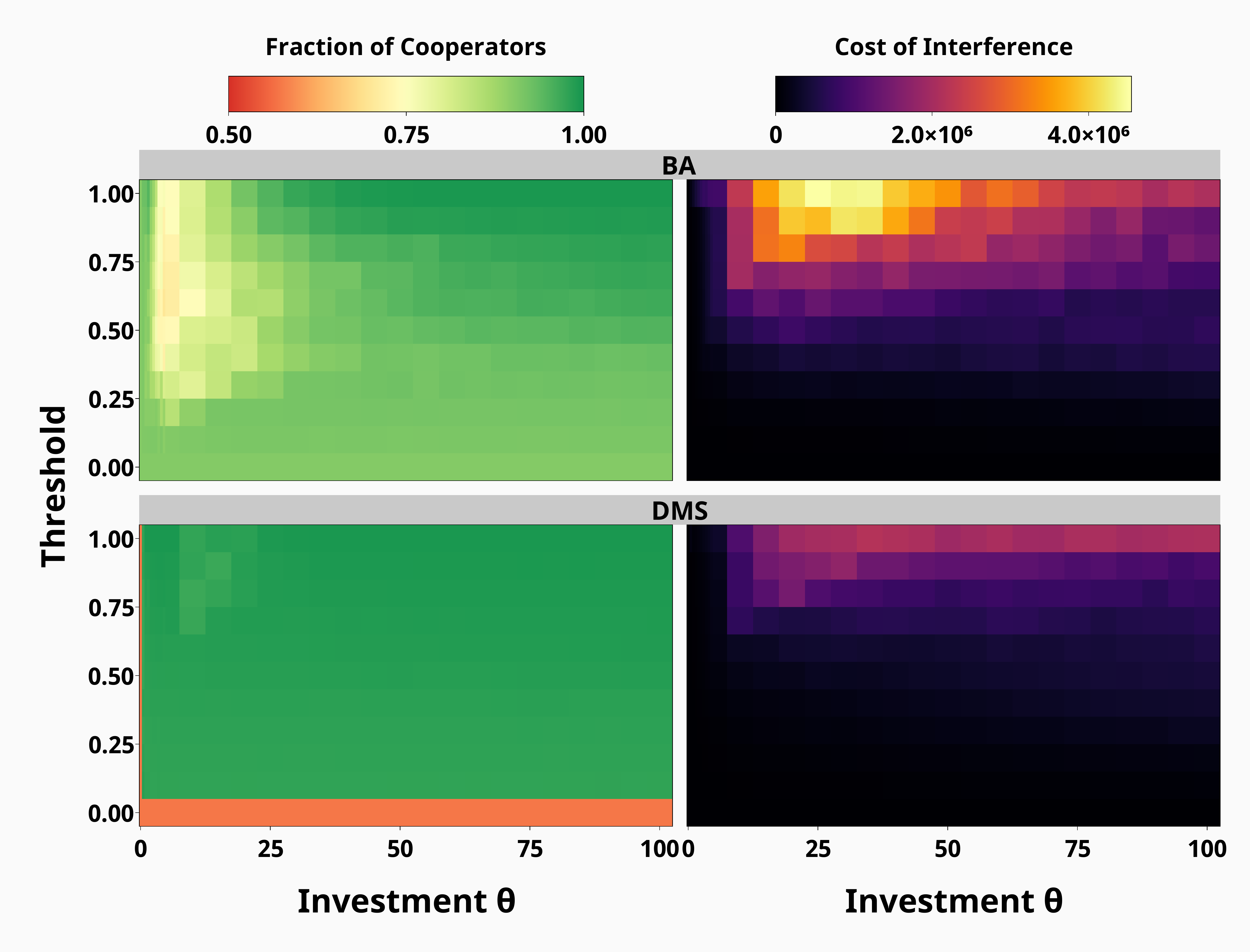}
\caption[Fraction of cooperation and total cost for population-based (POP) interference]{\textbf{Fraction of cooperation and total cost for population-based (POP) interference,} using deterministic update. Parameters: $b = 1.8; \ N = 5000$.}
\label{fig:pd-pop}
\end{figure}

Highly clustered networks often have the most room to improve by receiving endowments (See Figure \ref{fig:pd-pareto-det}). The initial distribution of players in the hubs of the network often determines whether the direction towards which the population will converge. Often, a small nudge can steer the population towards a desirable outcome (see Figure \ref{fig:pd-evol}). Moreover, this can easily be accomplished through a variety of disparate investment paradigms. For instance, metrics on the overall population (POP) can be used to guarantee maximal cooperation regardless of how the endowments are distributed (See Figure \ref{fig:pd-pop}). With the reduction in the complexity of designing an effective scheme, we look towards cost and ways to reduce overspending. Overeager endowments can lead to total costs several orders of magnitude larger than those applied as a last resort. Indeed, even very small endowments applied to few surviving cooperators can \textit{jumpstart} the formation of clusters resilient to invasion. Increasing the threshold for investment guarantees that more cooperators will be eligible for the endowments, thus exacerbating spending. Lowering this threshold guarantees that interference will only be triggered if desperately required. Investing in every cooperator as a last resort ensures pro-sociality. 

\begin{figure}[t]
\centering
\includegraphics[width=\linewidth]{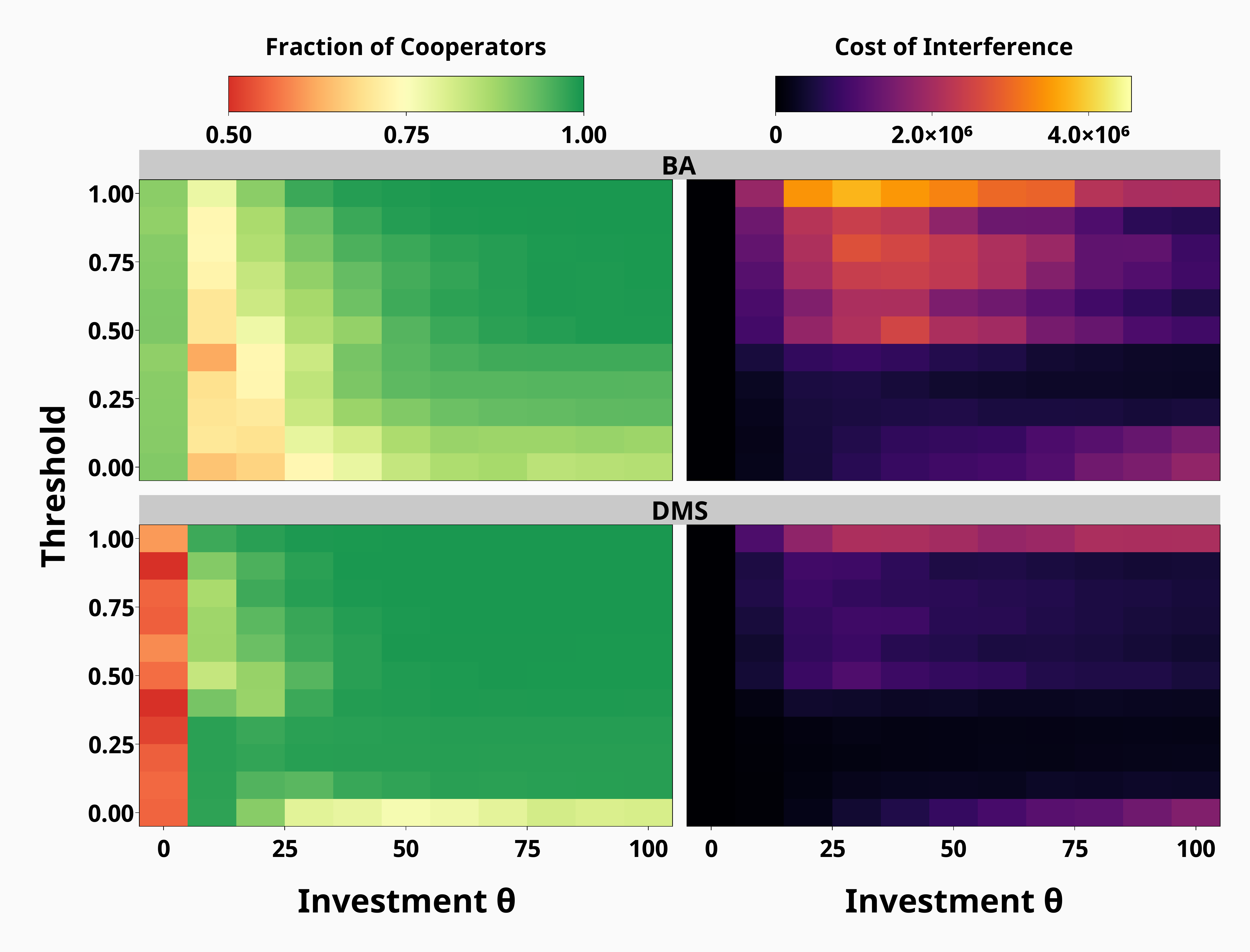}
\caption[Fraction of cooperation and total cost for local neighbourhood information (NEB) interference]{\textbf{Fraction of cooperation and total cost for local neighbourhood information (NEB) interference,} using deterministic update. Parameters: $b = 1.8; \ N = 5000$.}
\label{fig:pd-neb}
\end{figure}

Moreover, local observations can be used to ensure positive outcomes following a variety of pathways (see Figure \ref{fig:pd-neb}). In this case, an external decision maker must target a range of intermediate values for the threshold. Previous results on structured populations showed that investing in cooperator neighbourhoods with exactly one defector was the optimal way of fostering cooperation \citep{han2018ijcai}. In contrast, our findings suggest that the opposite is true for heterogeneous settings. Indeed, the least expensive routes towards cooperation are those with low or intermediate thresholds, suggesting that investors should focus their attention on ensuring only the survival of cooperators who are in danger of turning. For highly clustered networks, little investment is needed, and provided the threshold is not exceedingly low, maximal cooperation can be reached in any configuration, without unnecessary expenditure. Lowly clustered networks, on the other hand, require much more deliberate endowments to benefit from investment, with the added risk of causing cooperators to fall victim to exploitation as discussed previously in Figure  \ref{fig:pd-neb}. 

\subsection{Heterogeneity and network characteristics play a key role in the design of effective investment mechanisms}

\begin{figure}[t]
\centering
\includegraphics[width=\linewidth]{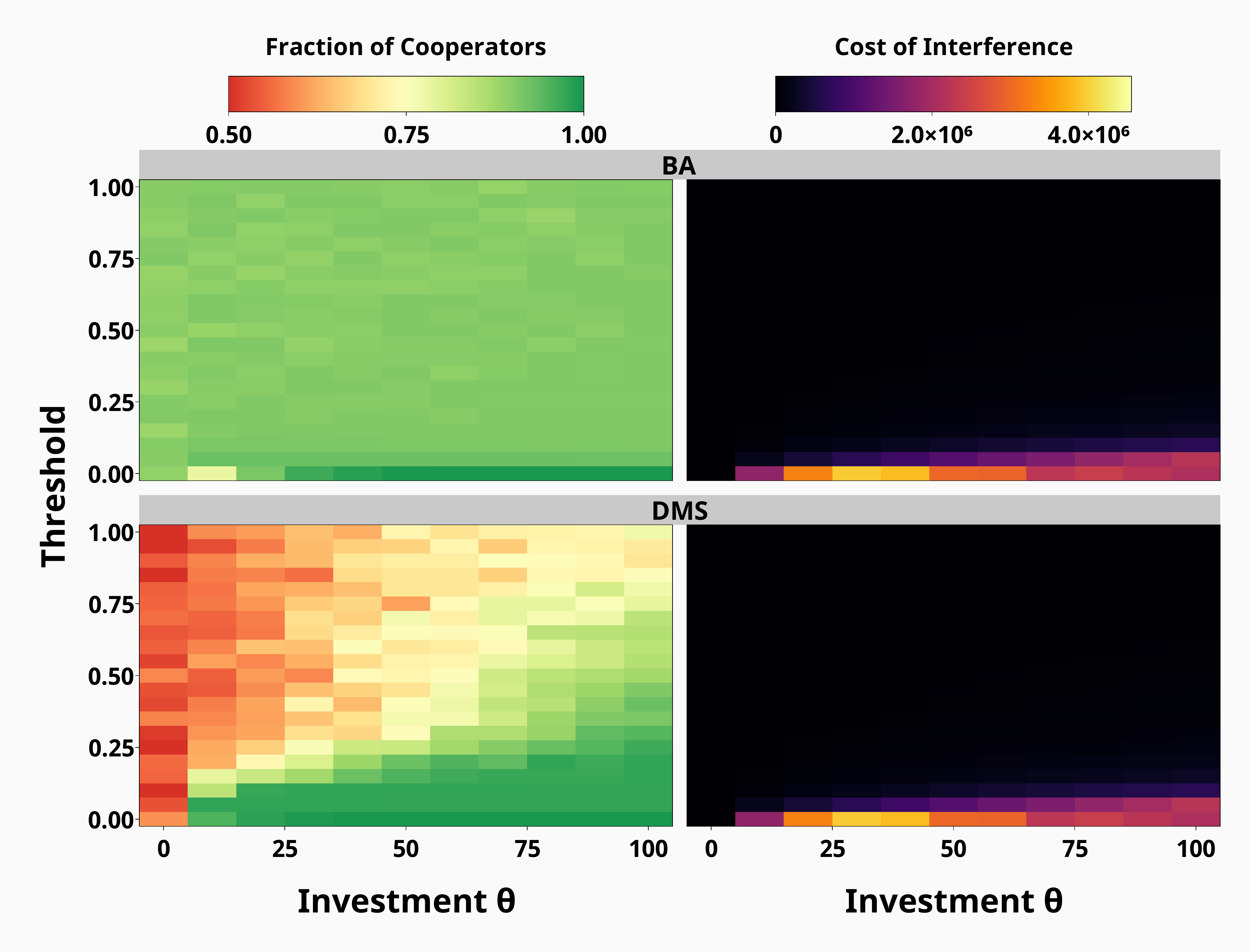}
\caption[Fraction of cooperation and total cost for node influence-based (NI) interference]{\textbf{Fraction of cooperation and total cost for node influence-based (NI) interference,} using deterministic update. Parameters: $b = 1.8; \ N = 5000$.}
\label{fig:pd-ni}
\end{figure}

Assuming that information about a node's influence can be easily gleamed by an external decision maker, this can provide a partial solution to reducing the risk of deleterious interference. Although comparatively costly, this mechanism has the benefit of never succumbing to the exploitation of cooperators (see Figure \ref{fig:pd-ni}). Notwithstanding, the very nature of influential nodes in scale-free networks (i.e. power-law degree distribution) implies only exceedingly large endowments are sufficient to sway them. However, the number of cooperators who are eligible for investment is also small; on account of this, overall spending does not scale predictably with the endowment amount. We have previously mentioned that there exist some costs associated with information gathering, which we do not model or measure here. Hence, the assumption that information about influence is readily available suggests this method could prevail in respect to real-world budgeting.

\begin{figure}[t]
\centering
\includegraphics[width=\linewidth]{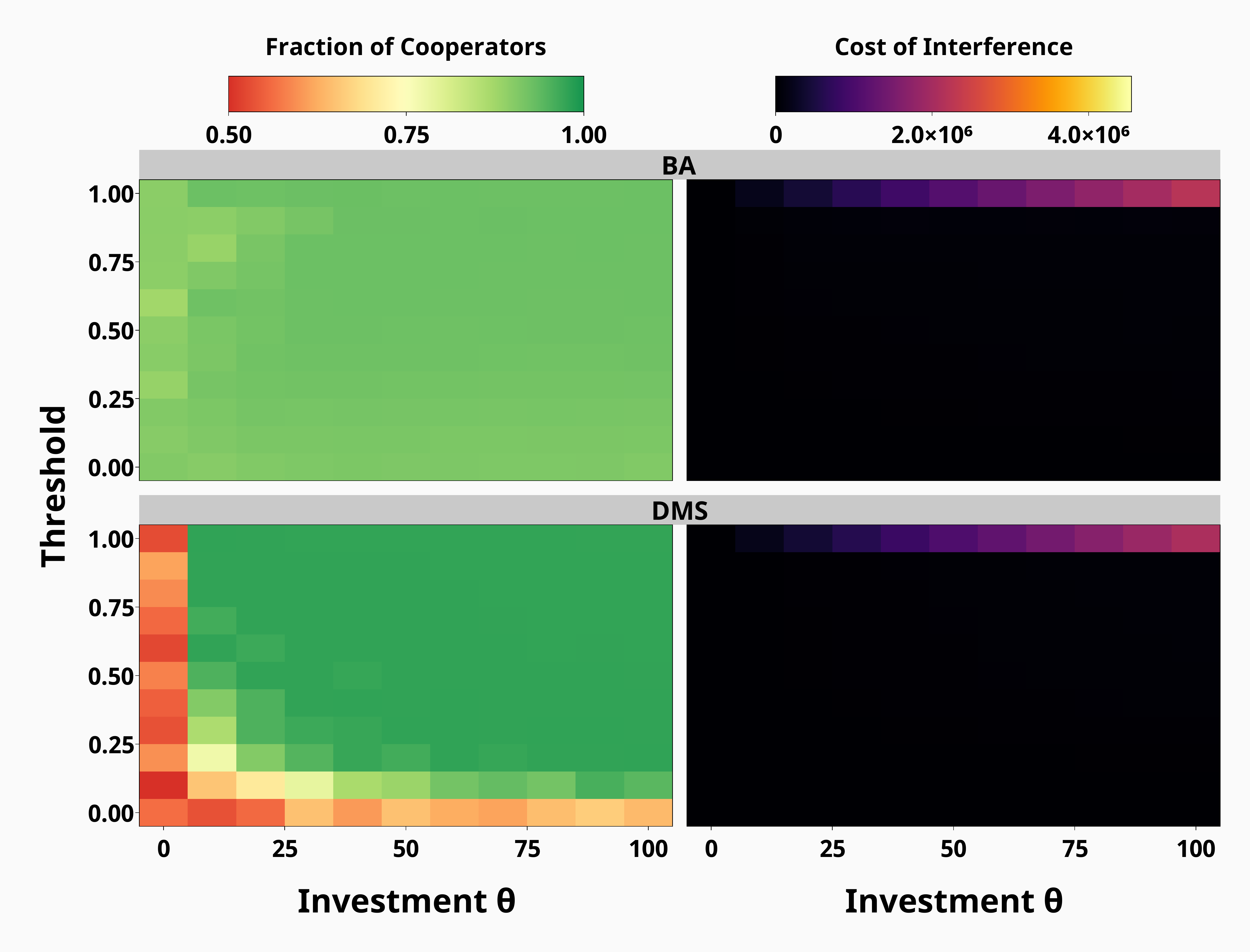}
\caption[Fraction of cooperation and total cost for a mixed interference scheme (NEB and NI)]{\textbf{Fraction of cooperation and total cost for a mixed interference scheme (NEB and NI),} using deterministic update. We fix $c_I = 0.05$, avoiding investing into the least connected nodes (bottom $5 \%$). Parameters: $b = 1.8; \ N = 5000$.}
\label{fig:pd-nebni}
\end{figure}

We propose that combining several interference mechanisms can be an effective way of reducing spending while avoiding the pitfalls of pernicious investment. For instance, we might consider taking into account an agent's influence as well as local observations. In Figure \ref{fig:pd-nebni}, we explore this possibility, avoiding the least connected nodes (i.e. not investing in the bottom $5 \%$ of nodes in respect to degree centrality), and show that this reduces spending compared to either of the two interference schemes taken individually. These results suggest that hubs play an important role in the emergence of cooperation in highly clustered networks, but that they cannot be effectively used to improve outcomes in their lowly clustered counterparts. Nevertheless, this integrated approach to interference eliminates the possibility of investment being detrimental to cooperation. 

We note this conundrum between the two types of heterogeneous networks. Lowly clustered networks have little to benefit from investment, and much to lose if the external investor is negligent in their distribution of endowments. On the other hand, highly clustered networks have much to gain and little to lose, readily responding positively to any tactic, overspending being the only matter of discontent. As investment in the greater context of heterogeneous interactions is not trivial, it would therefore be prudent to first collect as much data on the nature of the network before deciding to distribute endowments. Uncertainty about social diversity or clustering carries the additional risk of selecting an improper policy of designing incentive schemes. 

\subsection{Stochastic imitation increases the risk of exploitation}

\begin{figure}[t]
\centering
\includegraphics[width=\linewidth]{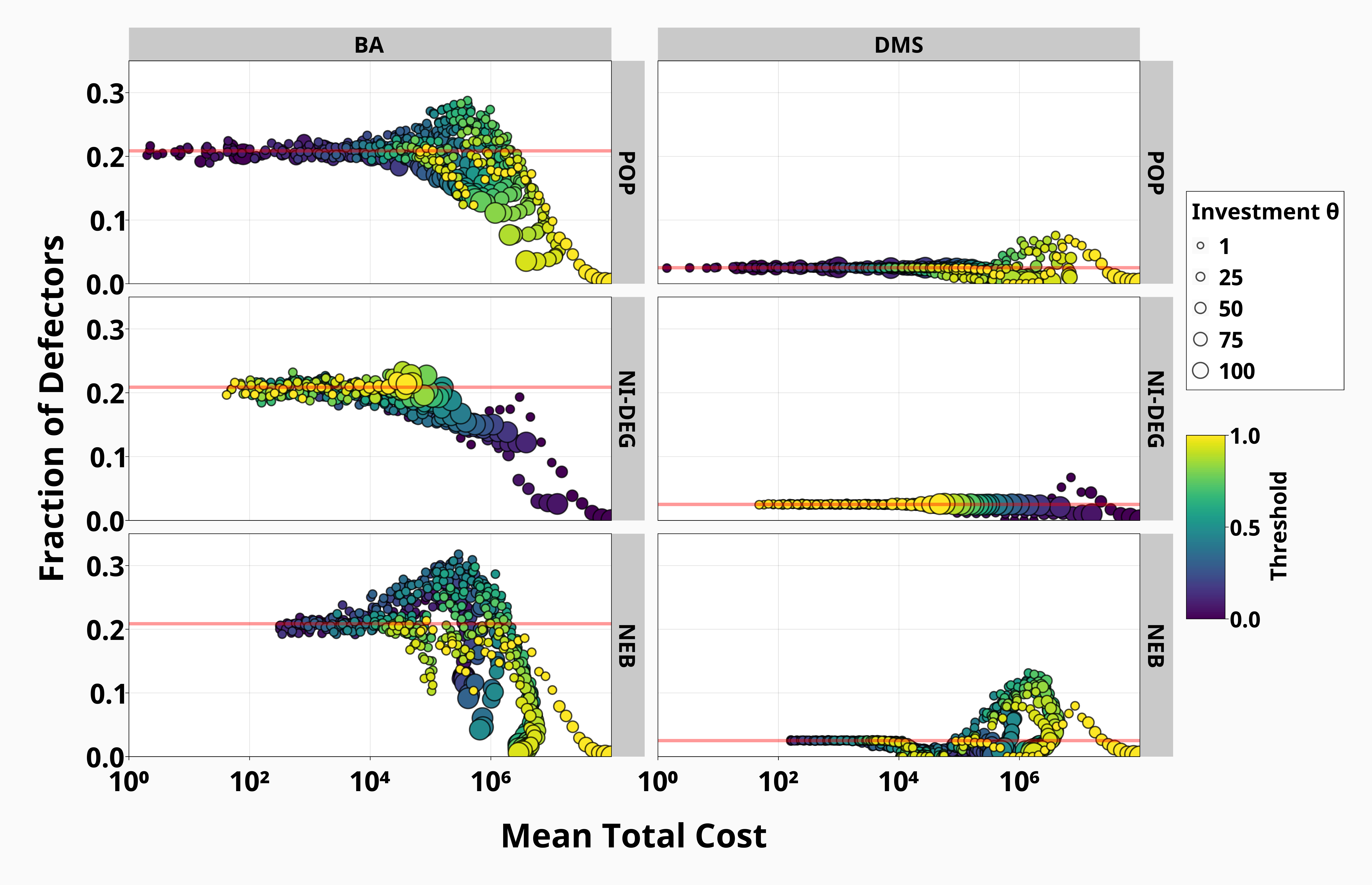}
\caption[Proportion of defectors as a function of the mean total cost for each scheme (stochastic update)]{\textbf{Proportion of defectors as a function of the mean total cost for each scheme (stochastic update)}. The markers' size is determined by the individual investment $\theta$ (grouped to the nearest value), whereas the colour indicates the threshold. Points near the origin indicate the optimal solutions. The horizontal red lines indicate the baseline level of defection in the absence of rewards for either network type. Parameters: $b = 1.8; \ N = 2000; \ k = 0.1$.}
\label{fig:pd-pareto-stochastic}
\end{figure}

\begin{figure}[t]
\centering
\includegraphics[width=\linewidth]{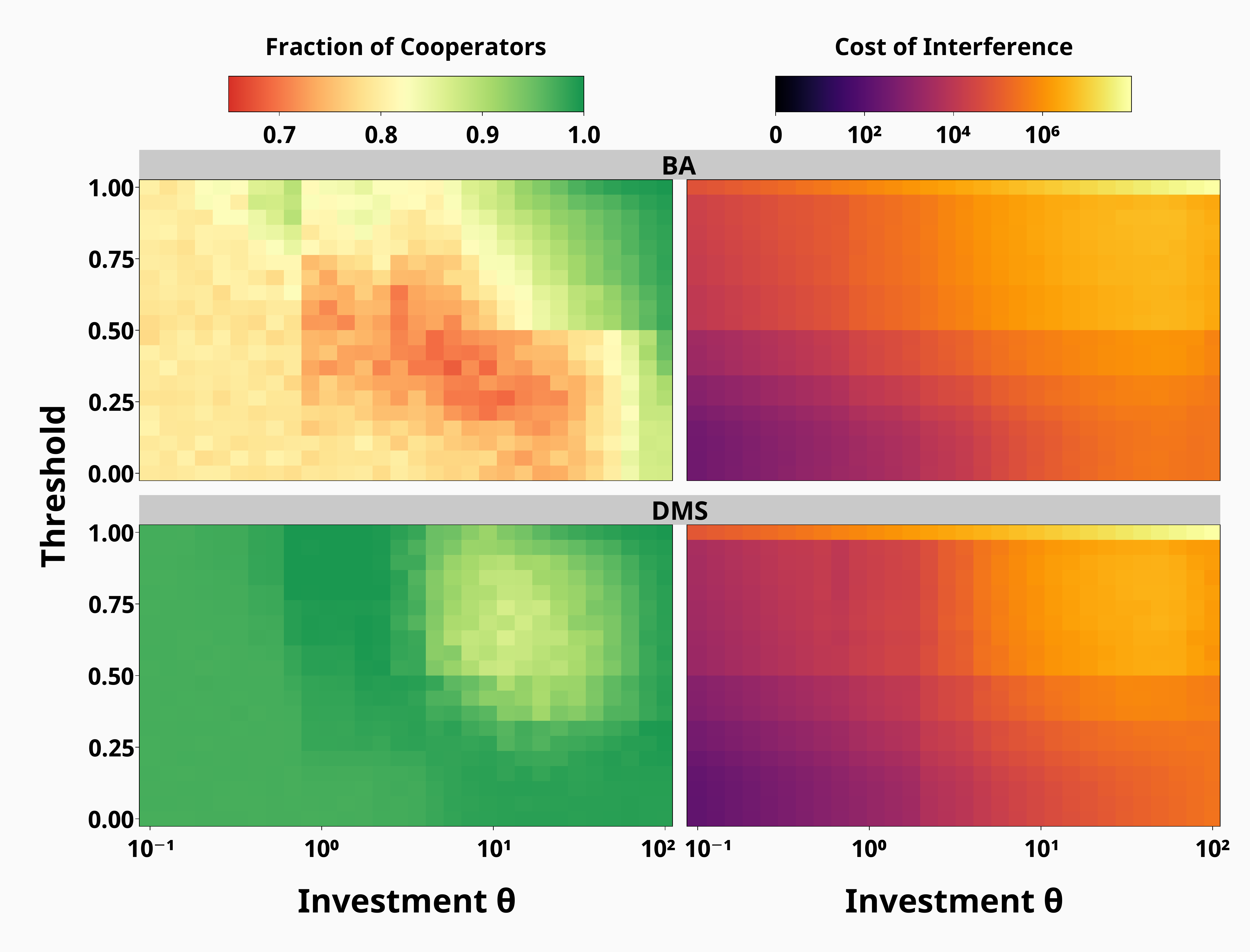}
\caption[Fraction of cooperation and total cost for local neighbourhood information (NEB) interference (stochastic update)]{\textbf{Fraction of cooperation and total cost for local neighbourhood information (NEB) interference,} using stochastic update. Parameters: $b = 1.8; \ N = 2000; \ k = 0.1$.}
\label{fig:pd-neb-st}
\end{figure}

Previously, we had shown that careless rewards might lead to an increase in defectors when interfering in BA networks under a deterministic update paradigm (see Subsection \ref{badrewards}). Following a transition towards a more realistic, stochastic update rule \citep{traulsen2006stochastic}, we observe a very similar phenomenon and moreover, find that it is no longer limited to lowly clustered scale-free networks (see Figure \ref{fig:pd-pareto-stochastic}). Indeed, investing in DMS networks should be approached with the same due diligence as BA networks, and insufficient endowments often lead to the exploitation of cooperators. Relying solely on local information is most prone to damaging cooperation, in spite of the level of complexity associated with this scheme and the amount of information required to enforce it (See Figure \ref{fig:pd-neb-st}).

\begin{figure}
\centering
\includegraphics[width=\linewidth]{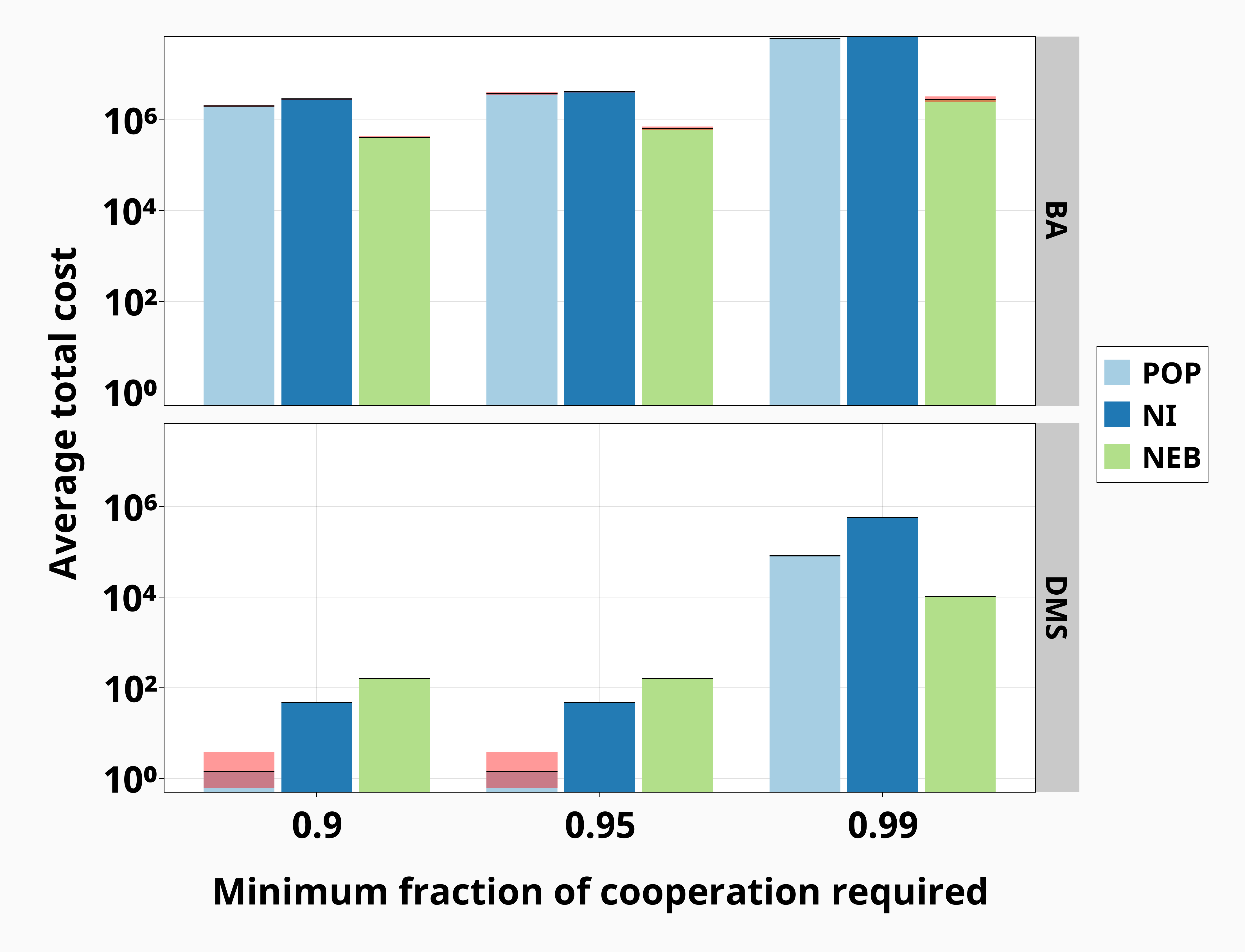}
\caption[Mean total costs for the most efficient combinations of threshold and investment amount (stochastic update)]{\textbf{Mean total costs for the most efficient combinations of threshold and investment amount} $\theta$, using stochastic update. We intentionally avoid configurations in which no endowments are distributed, and select the configurations with the least possible cost for each minimal fraction of cooperation required. Error bars in light red show the standard deviation across all replicates of a configuration.}
\label{fig:pd-boxplots-st}
\end{figure}
\begin{figure}
\centering
\includegraphics[width=\linewidth]{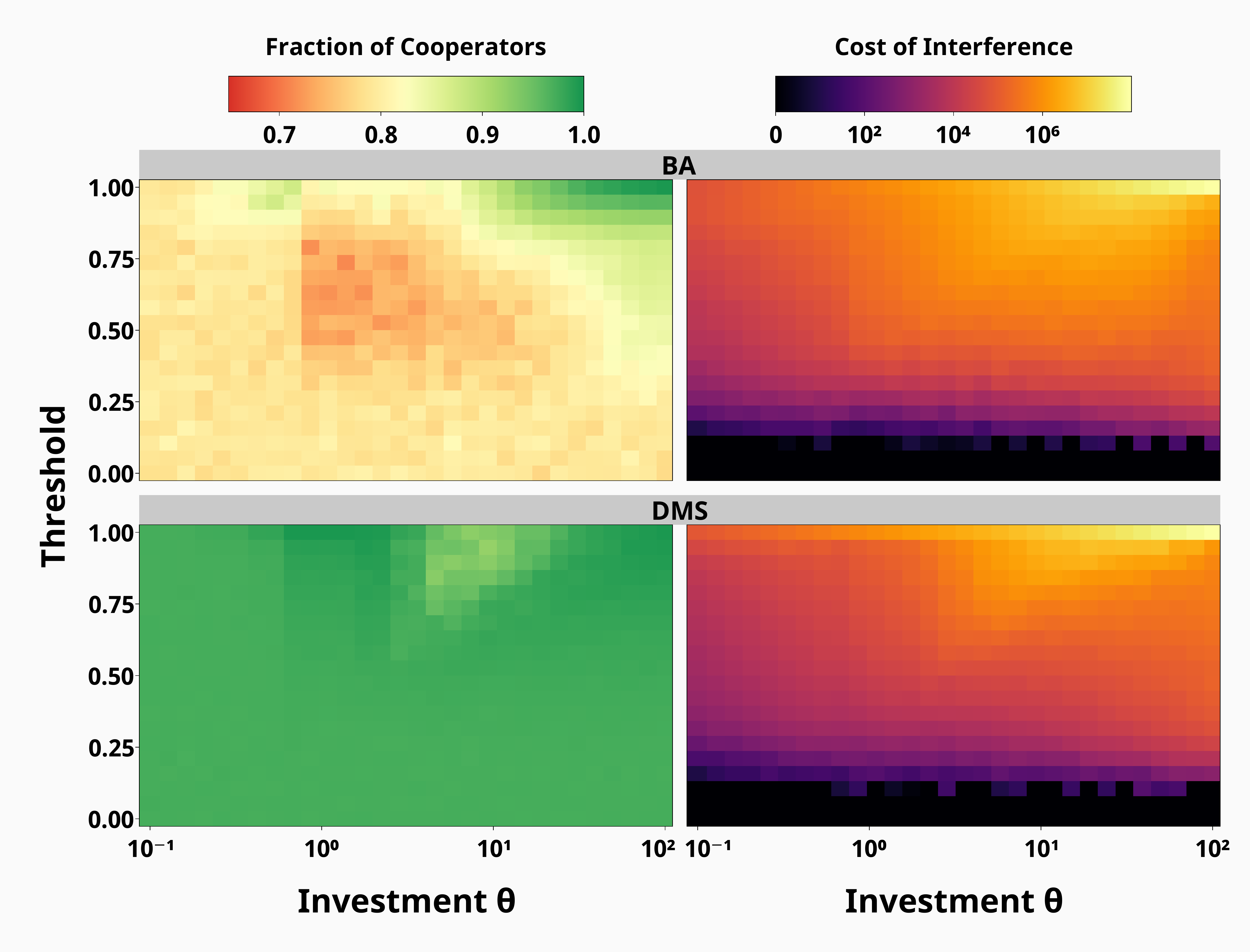}
\caption[Fraction of cooperation and total cost for population-based (POP) interference (stochastic update)]{\textbf{Fraction of cooperation and total cost for population-based (POP) interference,} using stochastic update. Parameters: $b = 1.8; \ N = 2000; \ k = 0.1$.}
\label{fig:pd-pop-st}
\end{figure}

Interestingly, stochastic imitation leads to a significant increase in baseline cooperation prior to interference in highly clustered networks, and conversely a decrease in cooperation in classical scale-free networks (see horizontal lines in Figure \ref{fig:pd-pareto-stochastic}). Nevertheless, the findings discussed in the sections above remain robust. Although the potential gains to be had shift, causing BA networks to benefit from investment more than their highly clustered counterparts, the previous findings still apply in this setting. For instance, we find that DMS networks readily respond to investment, and are not as prone to the pitfalls which befall BA networks. In Figure \ref{fig:pd-boxplots-st}, we show that the most efficient interference schemes are consistently one (or more) order(s) of magnitude less costly at promoting cooperation in highly clustered networks, regardless of what potential gains the external decision makers is aiming for. 

\subsection{Maximal cooperation gains require significant endowments. Cost-efficiency is a double-edged sword.}

\begin{figure}
\centering
\includegraphics[width=\linewidth]{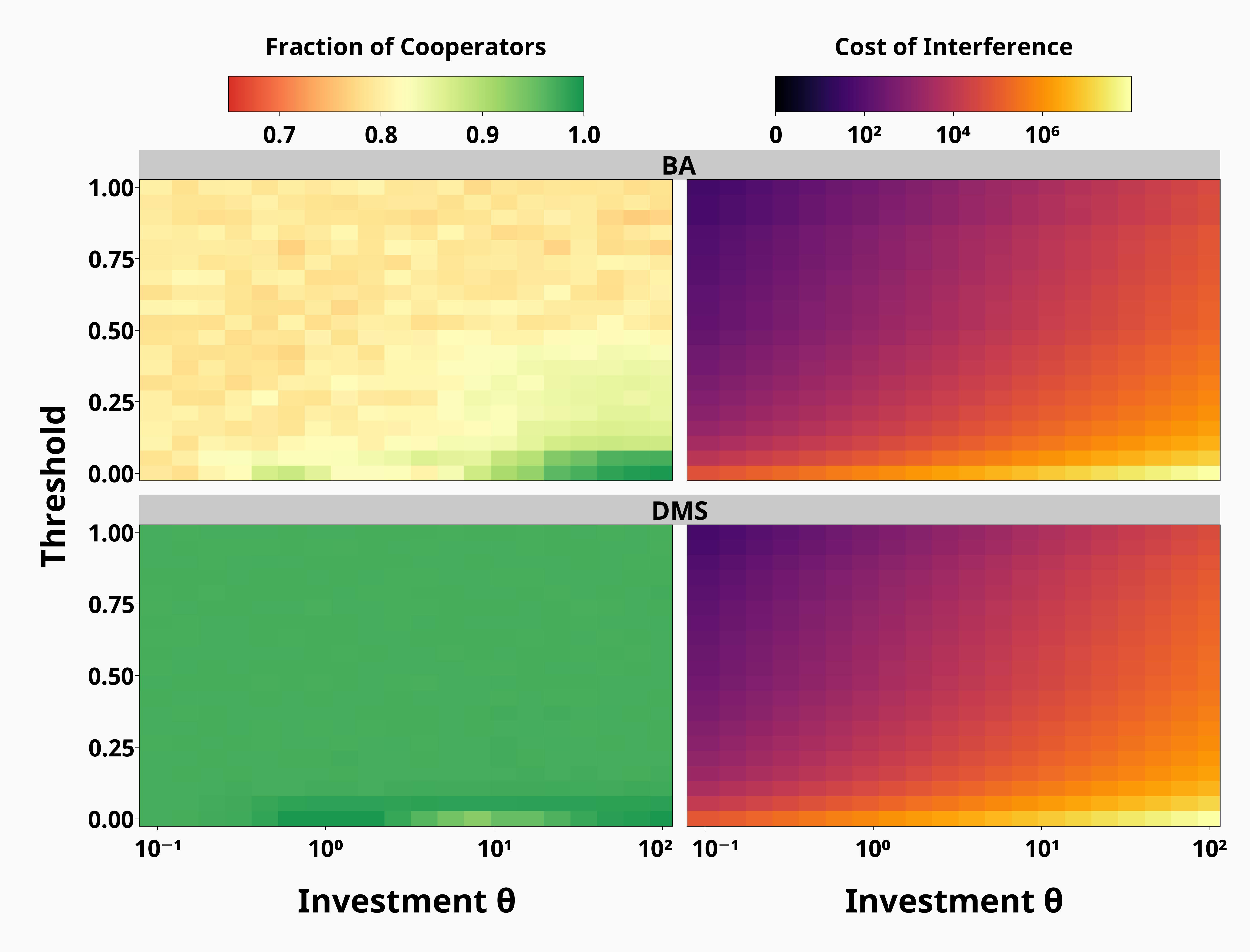}
\caption[Fraction of cooperation and total cost for node influence-based (NI) interference (stochastic update)]{\textbf{Fraction of cooperation and total cost for node influence-based (NI) interference,} using stochastic update. Parameters: $b = 1.8; \ N = 2000; \ k = 0.1$.}
\label{fig:pd-ni-st}
\end{figure}

Unlike homogeneous populations, heterogeneous interaction structures inherently provide a benefit to cooperators, something usually referred to as network reciprocity \citep{santos2006pnas, santos2008social}. In practice, this means investment does not lead to outcomes which differ significantly from the baseline. Furthermore, successful attempts at reaching a maximal level of cooperation (i.e. little to no defection) require a combination of large endowments and an investment scheme which can target individuals at all levels of the network (see Figure \ref{fig:pd-pareto-stochastic}). Using population level metrics generally fails to improve outcomes unless virtually every cooperator is targeted (see Figure \ref{fig:pd-pop-st}). Equivalently, relying on degree centrality (i.e. how influential a node is) necessitates an egalitarian distribution of endowments, which naturally increases costs (See Figure \ref{fig:pd-ni-st}).
Local information (NEB), while risky, also has the potential to best improve outcomes while reducing costs, and this remains true for both BA and DMS networks, if maximal cooperation gains are required (See Figure \ref{fig:pd-boxplots-st}). Once again, we intuit the importance of acquiring detailed observations of information about the agents. This approach is a double-edged sword; it is simultaneously the optimal solution, as well as the most prone to errors in decision-making, leading the population to either the most perceived gains or the least (See Figure \ref{fig:pd-neb-st}). This seems to be another dilemma. Investing is risky, and it is likely for endowments to be ineffective or even produce negative results, but only significant sums of capital are likely to lead to desirable outcomes. Social diversity complicates this further, as there exists a great degree of inequality between individuals, and potential errors in decision-making make investment precarious.

\section{Discussion}

In summary, we have studied how optimally  an external decision maker could incentivise a population of autonomous agents facing a cooperative dilemma to fulfil a coveted collective state. We build on a previous account which identified the most effective mechanisms to foster cooperative scenarios in spatially distributed systems in regular graph structured populations of agents, but instead we consider two popular models of scale-free networks of contacts. In particular, we examined  if the insights set out in the context of regular graphs remain applicable to heterogeneous models, as well as explore an additional avenue of interference enabled by the variance in node connectivity. To address these issues, we combined an evolutionary game theoretic model with several incentive mechanisms in two types of pre-generated networks characterised by preferential attachment, with different clustering coefficients. We argue that this problem cannot be solved trivially and we show that transitivity (i.e. the global clustering coefficient) should be the driving force behind the choice of an interference mechanism in promoting cooperation in heterogeneous network structures, as well as its application. 

In this work, we introduce several incentive mechanisms which are defined formally and mathematically. We note that they do not have to be defined as such, and in fact have many real-life counterparts which are often employed by institutions and investors. For instance, POP-based metrics describe cooperation observed at a global scale. If we consider the Great Recession, or the recent COVID-19 pandemic, an institution might only need to look at the overall state of the economy, or the spread of an infection, before deciding that action is required. Neighbourhood-based metrics represent local schemes, which are almost ubiquitous when considering social inequality. Whether it is housing schemes, incentives to stop smoking, homelessness, education, etc., local governments often decide to invest based on the level of economic and social deprivation in a specific area, and that is precisely what we have tried to capture with NEB-based schemes. Finally, we have looked at centrality (influence) metrics. If we consider social media, a company might wish to use influencers to market its products, or an institution might decide to specifically target someone in the public eye in order to increase the visibility of its incentives, whether they were positive or negative. The mechanisms we have chosen are by no means exhaustive choices, but they serve as a fundamental starting point to our discussion, and they are arguably the most common and most easily implementable mechanisms that we observe in the real world.

We find that impetuously rewarding cooperators can lead to cyclic patterns which damage cooperation in the long run, enabling the exploitation of cooperators to the benefit of defectors. We argue that detailed information gathering about the networks and agents prior to the distribution of endowments can prevent these mistakes. Using two social learning paradigms, we show the robustness of these findings and observe that clustering lowers the risk of deleterious investment, easing the strictness of distributing incentives. Moreover, we show that ignoring lowly connected individuals leads to unprofitable and even futile intervention irrespective of network transitivity. 

Our comparison between the two types of scale-free networks provides valuable insights regarding the importance of clustering in the outcome of cooperation. We find that a large clustering coefficient allows for successful, cost-effective interference, indeed even when partly disregarding a full comprehension of the population and its tendencies. Furthermore, transitivity lessens the burden on external investors, lowering the total cost required to enforce cooperation. These results are of particular interest, given that most SF networks portray high clustering, such as in the case of social ties where friends are likely to be friends of each other \citep{newman2018networks}. This scope encompasses heterogeneous scenarios inhibited by spatial constraints (e.g. in highly urbanised areas or even the allotment of rangelands such as pastures), where high clustering is also imposed. 

Transitioning towards a more realistic, stochastic imitation rule \citep{traulsen2006stochastic}, we measure a shift between the two network types, whereby lowly clustered networks prescribe a greater need for investment, and vice-versa. Maximal cooperation gains in either paradigm can generally be achieved using large individual endowments. Notwithstanding, highly clustered networks respond more readily to interference, and we provide several insights about ways in which cost could be reduced further.

An important question in models of institutional incentives   is that of setting up and maintaining the incentive budget. Considering who should contribute to the incentive budget is a social dilemma in itself, and addressing this second-order social dilemma has been identified as a  challenging research problem. Several solutions have been identified,  including pool incentives with second-order punishments \citep{sigmund2010social,perc2012sustainable}, democratic decisions \citep{hilbe2014democratic}, commitment formation \citep{han2015synergy,han2022institutional,sasaki2015commitment}, and hybrid incentives  \citep{chen2015first,gois2019reward}. This work does  not aim to address this issue, focusing instead on how to optimise the spending from a given  budget by exploiting network properties and information gathering. However, it would be interesting to study the co-evolutionary  institutional formation with different interference strategies and individual strategic behaviours.
Moreover, in this work, we do not consider the possibility of detecting the existence of a certain type of interference from an external party. In reality, individuals could be aware of active interference and react by changing their behaviour, either to become suitable candidates for reward or to avoid sanctions \citep{cimpeanu2020making}.

In summary, we have shown that it is crucial to investigate both the underlying network, and behavioural trends before distributing endowments. Previously, these findings were restricted to very particular combinations of parameters, but stochastic dynamics highlight the pitfalls of inconsiderate incentives in a variety of settings. As a result of such careless interference, defector communities grow before eventually collapsing due to their inability to reciprocate. We have identified several key metrics which can be used to mitigate these risks and ensure positive social outcomes. Our findings highlight the complexity of designing effective and cogent investment policies in socially diverse populations.



\end{document}